\documentclass[9pt]{book}

\usepackage[dvips]{graphicx,color}
\usepackage{makeidx,far,detector}

\usepackage{amsthm}
\usepackage{amstext}
\usepackage{amsmath}
\usepackage{amsbsy}
\usepackage{amssymb}
\usepackage{amsfonts}

\BookTitle{Far Detector in Korea for the J-PARC Neutrino Beam}
\CopyRight{\copyright 2007 by Universal Academy Press, Inc.}

\begin{document}
\pagenumbering{arabic}

\chapter{%
Summary of the 3rd International Workshop on a Far Detector in Korea for the J-PARC Beam}

\author{\raggedright \baselineskip=10pt%
{\bf T.\ Kajita,$^{1}$,
{\bf S.B.\ Kim,$^{2}$
and
A.\ Rubbia$^{3}$}\\ 
{\small \it %
(1) Research Center for Cosmic Neutrinos, Institute for Cosmic Ray Research (ICRR), 
and Institute for the Physics and Mathematics of the Universe (IPMU),
University of Tokyo, Kashiwa, Chiba 277-8582, Japan \\
(2) Department of Physics and Astronomy, Seoul National University, Seoul 151-742, Korea \\
(3) Institut f\"{u}r Teilchenphysik, ETHZ, CH-8093 Z\"{u}rich,
Switzerland
}
}}






     \baselineskip=10pt
     \parindent=10pt

\section{Introduction}

The 3rd International Workshop on a Far Detector in Korea for the J-PARC Neutrino Beam\footnote{http://www-rccn.icrr.u-tokyo.ac.jp/workshop/T2KK07/}\footnote{Workshop supported
by Japan-Korea Basic Scientific Cooperation Program of JSPS and KOSEF.} was held at the Hongo Campus of Tokyo University, 
Tokyo, Japan on Sep. 30th and October 1, 2007.  
Forty seven physicists from Japan and Korea, as well as Europe and USA, 
participated in the workshop and discussed the physics opportunities offered by the J-PARC
conventional neutrino beam detected by a new large underground neutrino detector in Korea.
In this paper, we highlight some of the most relevant findings of the workshop.

The present neutrino physics program in Korea, focused at a
reactor neutrino experiment RENO, and the state of the J-PARC neutrino
beam, presently under construction at Tokai (Japan)  for the T2K experiment,
were reviewed at the workshop. 
Future long baseline neutrino experiments have the task to complete the present knowledge on 
the mixing parameters, possibly including the CP violating phase $\delta$.
The measurements in RENO and T2K could indicate a non-vanishing  $\theta_{13}$ angle,
thereby ascertaining the $3\times 3$ nature of the lepton flavor mixing matrix. They may conclude that the 
simultaneous determination of the neutrino mass hierarchy and the CP 
violating phase is possible at a next generation long baseline experiment coupled
to a high intensity conventional neutrino beam.

The idea of a long baseline neutrino experiment from Japan to Korea as a future extension of  the J-PARC neutrino beam
program beyond T2K was debated, following the discussions held at the 1st and 2nd workshops
of this series. One of the specific purpose of this 3rd workshop was to further 
uncover the physics potential of the detector in Korea, to discuss the various issues related to detector technologies,
to address in more details systematic errors affecting the measurements,
to investigate possible synergies between RENO and the potential long-baseline program, and to globally
consider all physics opportunities offered by such an experiment.

Several options for the construction of a new large underground neutrino detector in Korea were addressed.
Many different parameters such as location, depth, off-axis angle, detector mass, or
detector technology, etc. can in principle be optimized to best detect the J-PARC neutrino beam.
In particular, the choice of the off-axis angle allows to optimize the neutrino beam shape, the smaller
off-axis angles yielding a broader and more energetic neutrino spectrum, while the larger angles corresponding to narrow band beams at
given neutrino energies. The discussion primarily centered on two categories of detector site locations; one is the site nearest to the 
on-axis beam trajectory which would receive a wide-band beam extending to a few GeV, and the other at 2.5$^o$ off-axis (OA)
for receiving a sub-GeV neutrino beam with the identical energy spectrum as in Kamioka. 

Water Cherenkov and liquid Argon detectors were discussed as candidates for detector technologies. 
The ``baseline setup'' assumed a very large deep underground Water Cherenkov imaging detector
of about 300~kton fiducial volume located in Korea at the same off-axis angle as Super-Kamiokande, complementing
another new similarly large detector at a new site in Kamioka. Similar configurations but with
different off-axis angles were considered, paying attention to background events, in particular for smaller OA angles.
A large 100~kton liquid Argon Time Projection Chamber at an OA1.0 in Korea was also considered, offering similar
and complementary physics performance. 

Several possible sites for the detector were mentioned, many of these with significant overburden, this latter
condition being most relevant for the study of atmospheric and supernovae neutrinos, and for the search of proton decay.
The Korean geological conditions are favorable, however more information for deep geological conditions
are missing at this stage. A more detailed geotechnical characterization will be needed to reduce potential risks for construction.

Overall, it was recognized through the workshop discussions that a long baseline experiment from Tokai to Korea
could give important information with which to understand the properties of neutrinos.
In particular, the neutrinos with long flight path in matter could be 
crucial to  determine the neutrino mass hierarchy, and 
lifting ambiguities in the measurement of the leptonic CP violation phase. Non-accelerator based
neutrino physics and the search for proton decay in a large underground detector would also address
fundamental questions of particle and astroparticle physics.

The results from the RENO and T2K experiments and more detailed studies are required to fully optimize the 
advantages of a Korean detector.  Development of detector technologies must continue and more detailed
investigations of the site are mandatory. It was decided to hold a forth workshop to further explore and update the 
opportunities of this very exciting physics program.

\section{The RENO reactor neutrino experiment}
An experiment, RENO ({\it R}eactor {\it E}xperiment for {\it N}eutrino
{\it O}scillation)~\cite{RENO}, is under construction to measure the smallest and 
unknown neutrino mixing angle ($\theta_{13}$) using
anti-neutrinos emitted from the Yonggwang nuclear power plant in Korea
with world-second largest thermal power output of 16.4 GW.
The experimental setup consists of two identical 15-ton Gadolinium loaded
liquid scintillator detectors located near and far from the reactor
array to measure the deviations from the inverse square distance law.
The near and far detectors are to be placed roughly 290~m and 1.4~km
from the center of the reactor array, respectively.
The experiment is planned to start data-taking in early 2010. An expected
number of observed anti-neutrino is roughly 5\,000 per day and roughly
100 per day in the near detector and far detector, respectively.
An estimated systematic uncertainty
associated with the measurement is less than 0.5\%. Based on three years of
data, it would be sensitive to measure the neutrino mixing angle in the
range of $\sin^2(2\theta_{13})>0.02$. This sensitivity is more than five times
better than the current limit obtained by CHOOZ~\cite{Apollonio:1999ae}.
 
\section{The T2K accelerator neutrino experiment}
The construction of J-PARC~\cite{ishida:JPARC}, the Japan Proton Accelerator Research 
Complex, a joint facility of High Energy Accelerator Research Organization (KEK) and Japan Atomic Energy 
Agency (JAEA) was in part inspired by the requirements of the T2K experiment~\cite{T2K},
the next generation long baseline experiment between J-PARC and Super-Kamiokande.
The primary motivation of T2K is 
to improve the sensitivity to the $\nu_{\mu} \rightarrow \nu_e$ conversion phenomenon 
in the atmospheric regime by about an order of magnitude 
compared to CHOOZ~\cite{Apollonio:1999ae}.

The J-PARC accelerator complex which includes
the 180~MeV LINAC, the 3~GeV Rapid Cycling Synchrotron (RCS) and the
30-50~GeV Main Ring Synchrotron (MR) is planned to be commissioned in 2008.
The J-PARC neutrino beam facility, under
construction for the T2K experiment, is foreseen to begin
operation in 2009~\cite{Ishida_Tokyo2007}.  
The final goal for the T2K experiment 
is to accumulate an integrated proton power on target of $0.75$~MW$\times5\times 10^7$ seconds.
Within a few years of run, critical information,
which will guide the future direction of the neutrino physics, 
will be obtained based on the data corresponding to about 
$1\div2$~MW $\times$ 10$^7$ seconds integrated proton power on target 
(roughly corresponding to a 3$\sigma$ discovery at 
sin$^2$2$\theta$$_{13}$$>$ 0.05 and 0.03, respectively)~\cite{NP08}. 

\section{Concept of the J-PARC-Korea long baseline neutrino program}
The J-PARC-Korea long baseline neutrino experiment is a natural continuation
of the physics addressed by RENO and T2K.
Let us start by describing  
the general concept of the J-PARC-Korea setup. 
We intend to make generic points which are valid without 
recourse to any specific setup. 

\begin{figure}[tb]
\begin{center}
\includegraphics[width=1.0\textwidth]{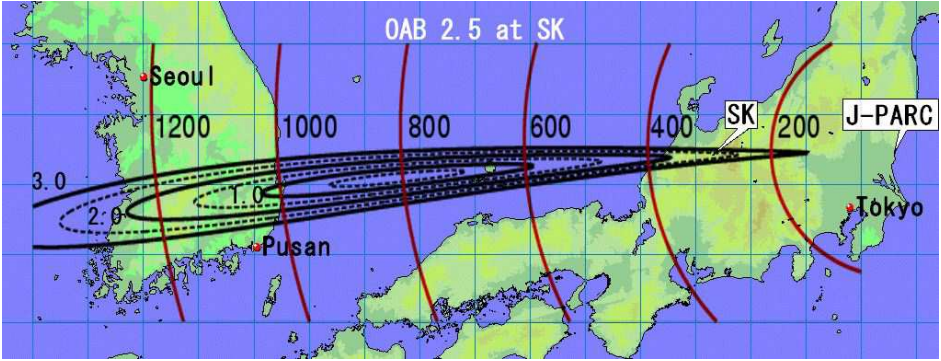}
\end{center}
\vglue -0.2cm
\caption{ Neutrino beam from J-PARC reappears in the Korean 
peninsula \cite{korea-map}. 
The solid lines with the numbers $n$ imply locations at which  
neutrino beam of $n$ degree off axis intersect with 
the earth surface (sea level). The red lines with the numbers 
denote the equi-distance curves from J-PARC in units of km. 
}
\label{korea}
\end{figure}

\begin{itemize}

\item 
The final goal for the T2K experiment is to 
reach an integrated intensity of $5 \times 10^{21}$ pots,
or equivalently a neutrino beam power of $\sim 0.75$~MW during 5 years.
There is a plan to further upgrade the accelerator complex to 
potentially provide an increased
beam power of 1.66~MW to the neutrino target~\cite{NP08}. This upgrade 
should in principle not require major modifications
in the beamline infrastructure which has been designed
up to 2~MW.
At this workshop, we assume an upgraded 4MW J-PARC beam created from 40
GeV protons, running $1.12 \times 10^7$ seconds per year was assumed. This is
equivalent to $28 \times 10^{21}$ POT per year.

\item

The neutrino beam from J-PARC, according to the current design, 
automatically reappears in the Korean peninsula. 
See Fig.~\ref{korea} which is taken from \cite{korea-map}. 
Therefore, 
it is a cost effective way to build a new experiment by having 
a far detector at an appropriate site in Korea, allowing in principle
simultaneous measurements at Kamioka (L=295~km) and
Korea (L$\approx$ 1000~km).

\item 
To determine the CP violating phase  and the neutrino mass
hierarchy, a powerful tool is to measure electron neutrino appearance
at both the first and second oscillation maximum. Two different
approaches are possible in order to make this measurement: 

(1) one
option is to have two detectors in the same beam (i.e. at the same off-axis angle), each of them
positioned mainly at one oscillation maximum, either the first or
second~\cite{Ishitsuka:2005qi}.
In this way many of the systematic errors can cancel or can be correlated with each other. 
The best situation could be achieved if both detectors were built in an
identical way.  This was defined as the ``baseline setup'';

(2) another approach, is to use a wide-band
energy beam, and measure electron neutrino appearance from both the
first and second maxima with the same
detector~\cite{Marciano:2001tz, Diwan_Tokyo2007}, by
realizing that if one can observe multiple oscillation maxima 
of neutrino oscillation, $\Delta m^2_{31} L / 2 E = (2n+1) \pi$ (n=0, 1, 2, ...), 
such measurement will have sensitivities to the mass hierarchy 
as well as CP violation phase in a single experiment. 

\item

By placing a detector somewhere in Korea at a baseline of about 1000~km, the experiment 
becomes sensitive to the matter effects in neutrino oscillation. 
It will give the experiment the ability of resolving the neutrino mass 
hierarchy, otherwise impossible at L=295~km.

\end{itemize}

\section{Determination of CP-violation phase and neutrino mass hierarchy}

\begin{figure}[htb]
\vglue -0.2cm
\begin{center}
\includegraphics[width=0.76\textwidth]{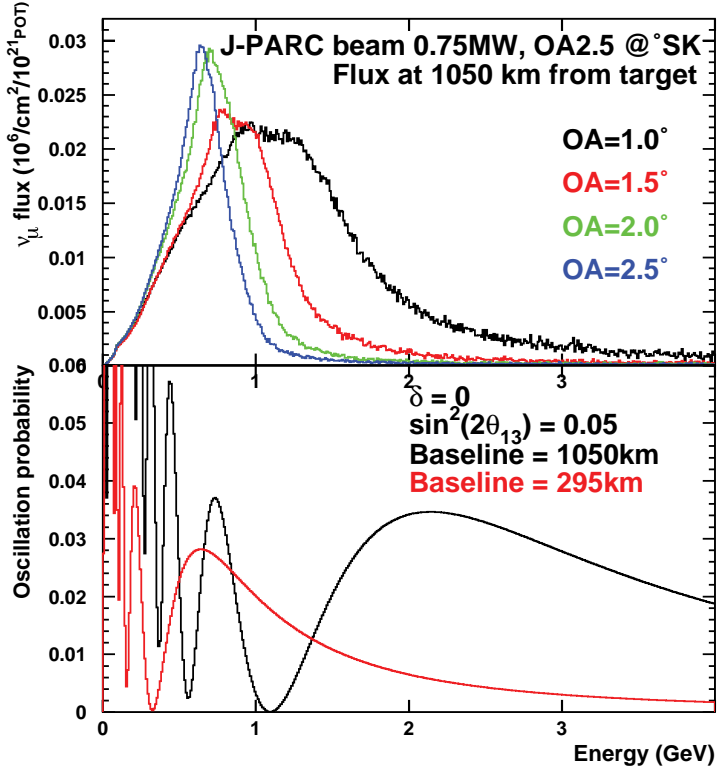}
\end{center}
\vglue -0.2cm
\caption{ Neutrino flux as a function of energy for several off-axis
    angle, and a 0.75MW beam at 1050km from the target.  For
    comparison, the $\nu_{\mu}$ $\rightarrow$ $\nu_e$ probability, for
    the two baseline considered in T2KK (295km and 1050km), for $\Delta
    m^2_{(21,31)}=7.3\times 10^{-5},2.5\times 10^{-3} eV^2$ and the
    other mixing angles at $\sin^{2}2\theta_{(12,23)}=0.86,1.0$. We
    assumed the earth density to be constant and to be equal to $2.8$
    $ g/cm^3$}
\label{fluxcomp}
\end{figure}

The neutrino beam spectrum in Korea will depend on the off-axis angle and
 on the exact geographical location chosen, because of the non-cylindrical
shape of the decay tunnel in the neutrino beam line~\cite{Rubbia-Meregaglia-Seoul2006}. 
When the upper side of the beam at $2^\circ$ to $ 3^\circ$ off-axis angle is observed at Super-Kamiokande,
the lower side of the same beam at $0.5^\circ$ to $3.0^\circ$ off-axis angle
can be observed in Korea.
As indicated in Fig.~\ref{korea} the 
J-PARC neutrino beam, to which the Kamioka detector is placed  
at 2.5 degrees off-axis, 
reappears in Korean peninsula as a beam with off-axis angle 
larger than 1 degree. 
Then, depending upon the off-axis angle chosen, a wide range of 
neutrino energy spectra becomes available~\cite{Rubbia-Meregaglia-Seoul2006},  
as exhibited in Fig.~\ref{fluxcomp}.

In the first published article\cite{Ishitsuka:2005qi}, the
off-axis angle of the Korean detector was assumed to be $2.5^{\circ}$.
The expected performance of this setup was recalled at this workshop~\cite{Kajita_Tokyo2007}.

It was argued that using a higher energy beam is better for the 
mass hierarchy determination due to the larger matter effect
\cite{Hagiwara, Okamura_KIAS2005}. 
However, experimentally, one expects higher background rate
in the sub-GeV energy range for the higher energy beam due to 
the larger amount of neutral current contamination. 
Therefore one has to estimate
the expected background carefully in order to compare the 
sensitivities of the low and high energy beam options.

\begin{figure}[htbp]
\begin{center}
\includegraphics[width=0.8\textwidth]{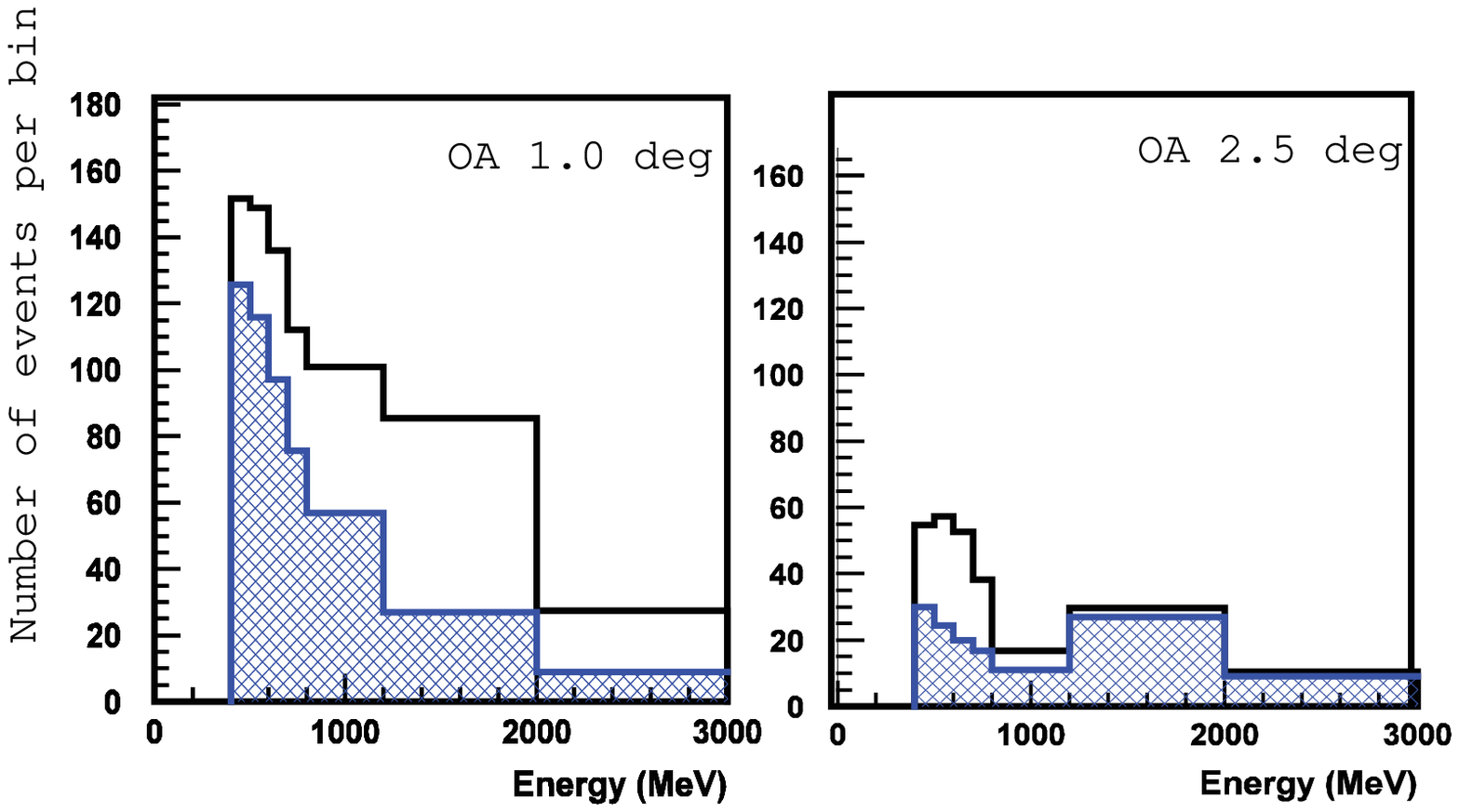}
\end{center}
\vglue -0.2cm
\caption{ Expected background (hatched regions) for the electron 
appearance search
for the Korean detectors with off-axis angles 1.0(left) and 2.5(right) degrees.
Also shown are the expected signal (solid histograms) over the background 
for $\sin ^2 2 \theta_{13}=0.1$ and $\delta = \pi/2$. }
\label{BG_angle-dep}
\end{figure}

The possibility of using wide-band beam for the detector in Korea 
with the background estimation was discussed
\cite{Dufour-Seoul2006,Rubbia-Seoul2006}. 
In Fig.~\ref{BG_angle-dep} the understanding of the  
signal and background events in a water Cherenkov detector
for various off-axis angle is 
presented \cite{Dufour-Seoul2006}.
One can recognize that there is an accumulation of background 
events at low energies which comes from high energy tail of the 
neutrino energy spectrum. 
This feature makes it highly nontrivial to reject background in an 
unambiguous way in water Cherenkov detectors.

\begin{figure}[tbp]
\begin{center}
{\hbox{\hspace{0.0in}
    \includegraphics[height=2.5in]{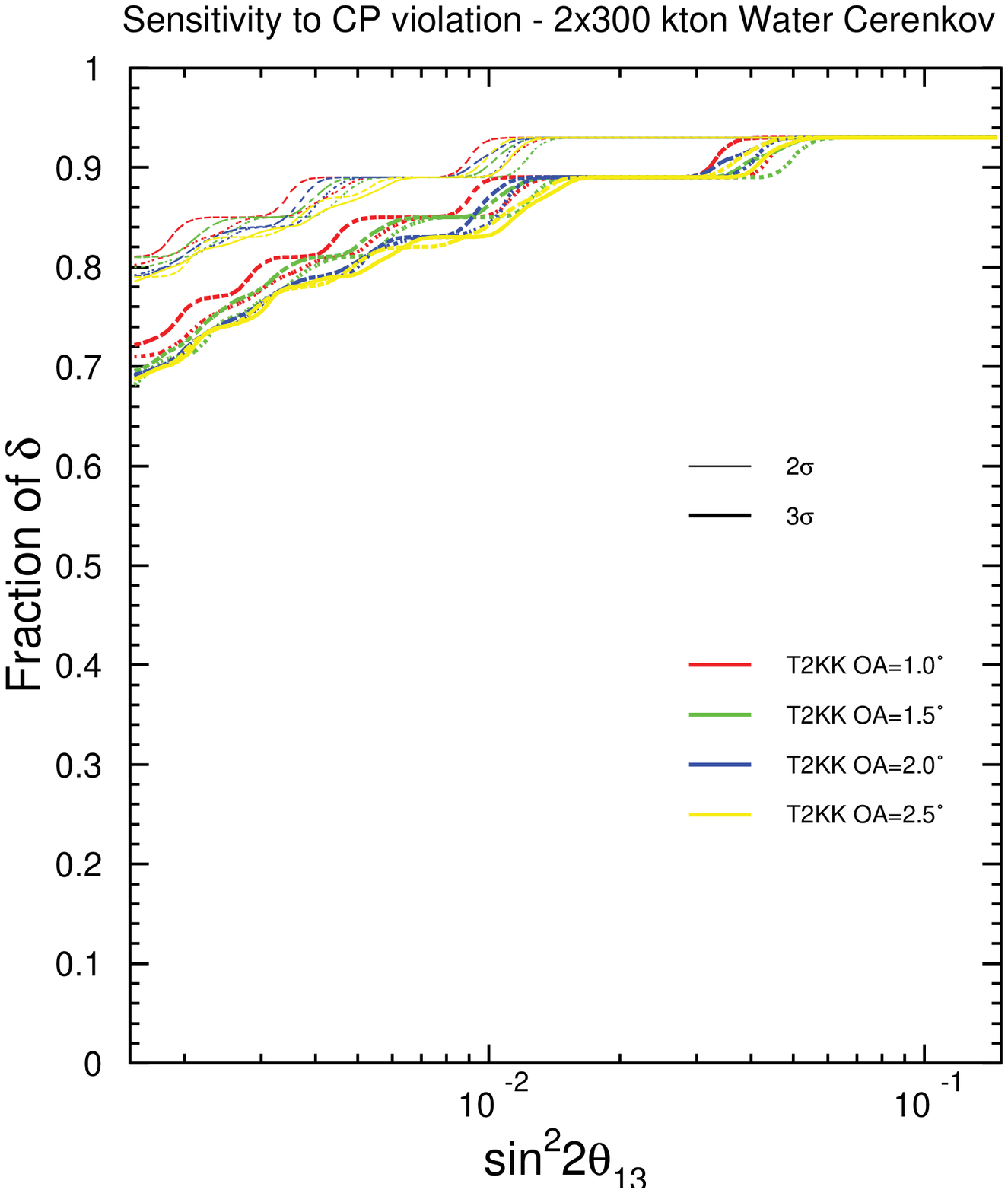}
\hspace{0.0in} \includegraphics[height=2.5in]{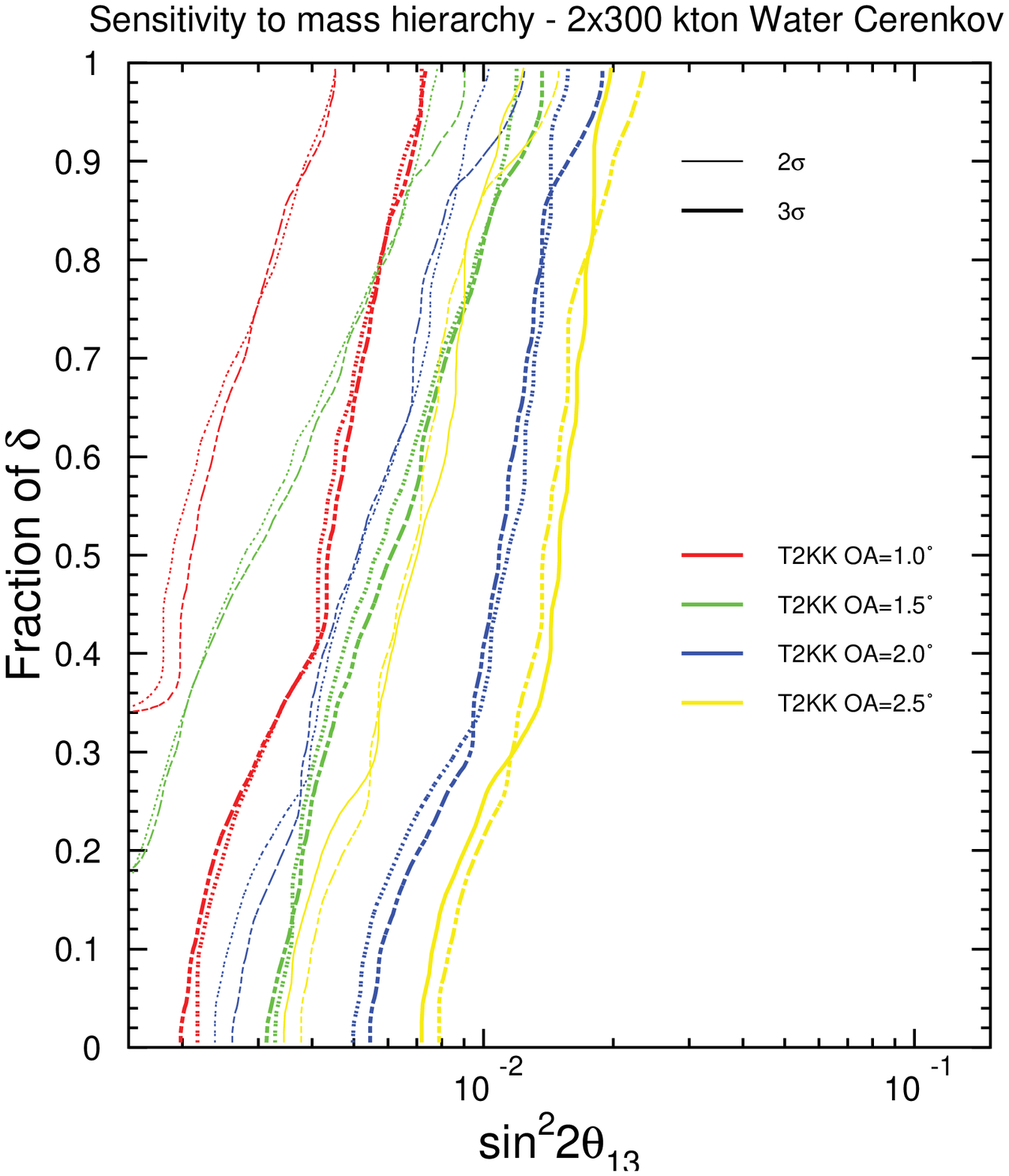}
}}
\end{center}
  \vspace{-1.5pc}
  \caption{ Sensitivities~\cite{Dufour_Tokyo2007} to CP violation (left) and mass hierarchy
    (right) for different values of the off-axis angle assuming $2\times 300$~kton fiducial volume
    Water Cerenkov detectors. (Other
    parameters: $\Delta m^2_{(21,31)}=7.3\times 10^{-5},2.5\times
    10^{-3} eV^2$ and the other mixing angles:
    $\sin^{2}2\theta_{(12,23)}=0.86,1.0$. We assumed the earth density
    to be constant and to be equal to $2.8$ $ g/cm^3$)}
\label{fraction_angle-dep}
\end{figure}

With the current understanding of the background one can examine 
if the near on-axis detector in Korea improves the sensitivities 
to the mass hierarchy and CP violation.  
In Fig.~\ref{fraction_angle-dep}, off-axis angle dependence of the 
sensitivity to the mass hierarchy  and CP violation
are shown \cite{Dufour_Tokyo2007}.
The sensitivity to the mass hierarchy resolution improves with the
decreasing off-axis angles (namely with the increasing beam energy)
while the CP violation results remain essentially intact. This is the most 
attractive feature of the near-on-axis option of the Korean detector.
The improved sensitivity to the mass hierarchy is indeed expected. 
The near on-axis Korean detector can cover the neutrino energy 
spectrum including the first and the second oscillation maxima with 
much higher event rate in the former region, i.e., at higher energies.
The matter effect is stronger in this region and hence the higher 
resolving power for the mass hierarchy. 
%

However, a cautionary remark for the
interpretation of the above results has been expressed~\cite{Kajita_Tokyo2007}. 
The statistical procedure used to produce Fig.~\ref{fraction_angle-dep} is identical with that 
used in the analysis of two identical detector case \cite{Ishitsuka:2005qi}, 
which means that most of the systematic errors are assumed to be
completely correlated between the two detectors. 
Therefore, a careful reanalysis is called for including more
realistic systematic errors~\cite{Huber_Tokyo2007,Okumura_Tokyo2007,Okamura_Tokyo2007}. 
It was also pointed out that a careful treatment of the Earth matter profile
is mandatory~\cite{Senda_Tokyo2007}.
Nonetheless, improvement of potential for the mass hierarchy 
determination is likely to survive in a proper treatment 
because of the physics arguments presented above. 

\begin{figure}[htbp]
\vglue -0.2cm
\begin{center}
\includegraphics[width=0.48\textwidth]{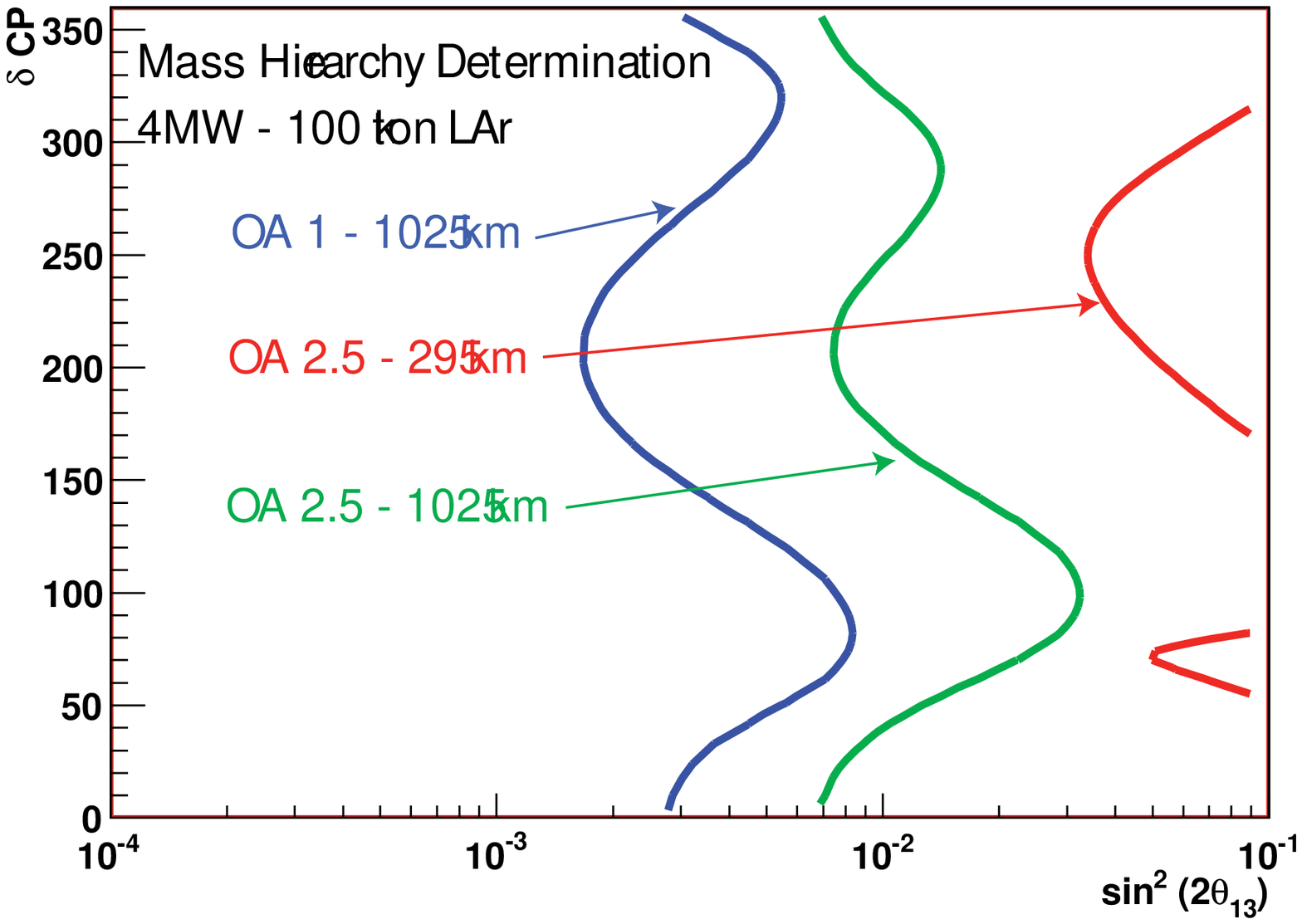}
\includegraphics[width=0.48\textwidth]{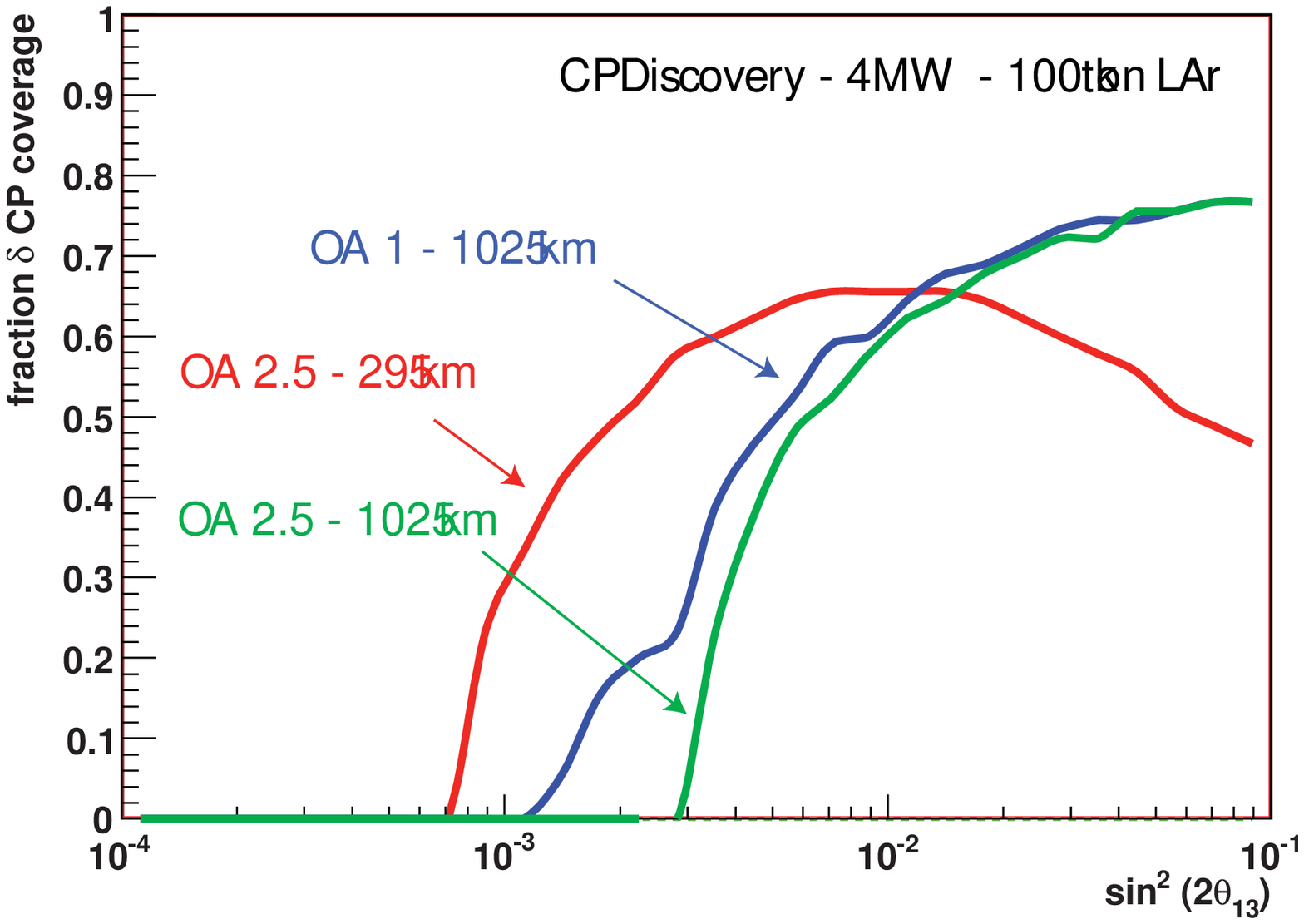}
\end{center}
\vglue -0.2cm
\caption{  The expected sensitivities~\cite{Rubbia_Tokyo2007}  to the mass hierarchy (left)
and CP violation (right) are presented in the form of the CP fraction  
assuming a single 100~kton LAr~TPC far detector in Korea.
The sensitivity to mass hierarchy is based on the neutrino beam only.
The dotted and solid lines are for sensitivities at 90\% and 
3$\sigma$ CL, respectively.
}
\label{LAr-sensitivity}
\end{figure}

Another option one could take for the high energy beam
is to use a much advanced detector technology to consider all events around the GeV region and above
(while the WC technology is essentially limited to quasi-elastic
events)  and simultaneously reduce the 
neutral current background as much as possible. One such 
example could be a very large Liquid Argon Time Projection Chamber~\cite{Rubbia_KIAS2005}.
The imaging properties and the good energy resolution
of the LAr TPC would allow studying
the broader band beam  with
the OA1 off-axis angle, covering more
features of the oscillation probability (e.g. first maxima, first minima,
second maxima, etc.)~\cite{Rubbia-Seoul2006}.
In such detectors, the rejection of the neutral current background
events will be achieved much more efficiently, reducing the background
events in the higher energy beam \cite{Rubbia_KIAS2005,Rubbia-Seoul2006}. 
Fig.~\ref{LAr-sensitivity} shows the expected sensitivity
of the liquid argon detector located at the 1.0 degree 
off-axis in Korea \cite{Rubbia_Tokyo2007}. 
It is clear that the sensitivity of the experiment
will be very high even for very small $\sin^2 2 \theta_{13}$
values.

In addition to the physics topics summarized so far, 
various other possibilities of physics measurements to be carried out with the 
Korean detector and the neutrino beam were discussed at the workshop~\cite{Ishihara_Tokyo2007,Ko_Tokyo2007,Kimura_Tokyo2007}. 

One such example was the sensitivity to the octant of 
$\theta_{23}$ based on data from RENO and a next generation underground
long-baseline neutrino experiment in Korea. It was
concluded that the combined results should show the 
sensitivity in the octant of $\theta_{23}$, if 
$\sin ^2 2 \theta_{13}$ is large ($> 0.05$)~\cite{Ishihara_Tokyo2007}.

Another interesting study was on the non-standard
neutrino interactions and properties. It is interesting 
to note that
the two detectors setup is powerful to constrain several 
non-standard nature of neutrinos~\cite{Ko_Tokyo2007}.  

\section{Detector technologies} 
The two detector technologies considered are a massive deep underground Water
Cherenkov imaging (WC) detector with a fiducial mass of 300-500 kton, and a fully
active finely grained liquid Argon time-projection-chamber (LAr TPC)
with a mass of $\sim$ 100 kton.

\subsection{Large Water Cherenkov Imaging detector}
Two generations of large water Cherenkov detectors at Kamioka (Kamiokande\cite{Koshiba:mw} and Super-Kamiokande\cite{Fukuda:2002uc})
have been very successful in research of neutrino physics with astrophysical sources. In addition,
the first long baseline neutrino oscillation experiment with accelerator-produced neutrinos, K2K~\cite{Ahn:2006zza},
has been conducted with Super-Kamiokande as far detector. 
Super-Kamiokande is composed of a tank of 50~kton of water (22.5~kton fiducial) which is
surrounded by 11146 20-inch phototubes immersed in the water. About $170~\gamma/cm$ are produced
by relativistic particles in water in the visible wavelength $350<\lambda<500~nm$. With 40\% PMT coverage
and a quantum efficiency of 20\%, this yields $\approx 14$ photoelectrons per cm or $\approx 7$~p.e. per MeV 
deposited.

There are good reasons to consider a third generation water Cherenkov detector with an
order of magnitude larger mass than Super-Kamiokande: a megaton Water Cherenkov detector will have a 
broad physics programme, including both
non-accelerator (proton decay, supernovae, ...) and accelerator physics.

Hyper-Kamiokande~\cite{Nakamura:2003hk} has been
proposed with about 1~Mton, or about 20 times as large as Super-Kamiokande, based
on a trade-off between physics reach and construction cost. 
Further scaling is limited by light propagation in water (scattering, absorption).
Although this order of magnitude
extrapolation in mass is often considered as straight-forward, a number of R\&D efforts including the site
selection are needed before designing the real detector. 
An important item for Hyper-Kamiokande is developments of new photo-detectors: with the
same photo-sensitive coverage as that of Super-Kamiokande, 
the total number of PMTs needed for Hyper-Kamiokande will be $\simeq 200000$. Possibilities
to have devices with higher quantum efficiency, better performance, and cheaper
cost are being pursued. 

\subsection{Liquid Argon Time Projection Chamber}
Among the many ideas developed around the use of liquid noble gases, the Liquid 
Argon Time Projection Chamber (LAr TPC) (See Ref.~\cite{t600paper}
and references therein)  certainly represented one of the most
challenging and appealing designs.
The LAr TPC  is
a powerful detector for uniform and high accuracy imaging of massive active volumes. 
It is based on the fact that in highly pure Argon, ionization tracks can be drifted
over distances of the order of meters. 
Imaging is provided by position-segmented electrodes at the end of the drift path, continuously recording the 
signals induced. $T_0$ is provided by the prompt scintillation light. 

One possible design for a detector with mass order of 100~kton, called GLACIER concept~\cite{Rubbia:2004tz},
was assumed at this workshop. 
The pros and cons of the LAr TPC, in particular in comparison to the Water
Cherenkov Imaging technique,  can be summarized as follows:

The liquid Argon TPC imaging should offer optimal conditions
to reconstruct the electron appearance signal in the energy
region of interest in the GeV range, while considerably 
suppressing
the NC background consisting of misidentified $\pi^0$'s. 
The signal efficiency
is expected to be higher to that of the WC detector, hence, the LAr TPC detector
could be smaller: the 100~kton detector considered here is
approximately twice the size of the
Super-Kamiokande detector.
In addition, a LAr TPC should allow operation at shallow
depth. The
constraints on the excavation and the related siting issues of the detector
should hence be reduced compared to WC.

The community has less experience with the LAr TPC technology 
than the WC; the largest detector
ever operated, the ICARUS T300, has a modular design which is not easily
extrapolated to the relevant masses. Significant R\&D and improvements
in the design are therefore
required in order to reach a scalable design which could offer a path for
a 100~kton mass facility in a cost effective way.

The procurement and underground
handling of large amounts of liquid Argon is more difficult 
than that for water, however, safe, surface or near-surface,
storage of very large amounts of cryogen (with volumes larger
than the ones considered here) has been achieved by the petrochemical industry; liquid Argon
is a natural by-product of air liquefaction which has large industrial and commercial applications
and can be in principle produced nearby any chosen location.

\section{Preliminary site study}
Throughout Korea, mountains are not high, rarely exceeding 1,200 meters, but they are found almost everywhere. The terrain is rugged and steep, and only near the west and southwest coasts are extensive flat alluvial or diluvial plains and more subdued rolling hilly lands. 
Several sites have been considered in a preliminary way: the preferable locations are either a mountain, a mine, or a tunnel.
The use of abandoned/closed mines might have some advantages, however, none seemed to satisfy the necessary conditions.
Large underground caverns in Korea in the range of 100000~$m^3$ and depths ranging from
150~m down to 350~m exist and are used for oil, LPG, food and water storage.
The general procedure for underground civil engineering construction foresees  (a) preliminary data collection 
(preliminary assessment, preliminary geotechnical characterization), (b) a feasibility study 
(engineering classification of rock mass, feasibility assessment of tunneling problems \& alternatives),
(c) a detailed site characterization (d) stability analyses  and (e)  final design and construction.
These have not yet been fully addressed.
In conclusion, mountains or shallow depth caverns seem to be most adequate solutions for siting the detector.
The Korean geological conditions are favorable, however more information for deep geological conditions
are missing at this stage. A more detailed geotechnical characterization will be needed to reduce potential risks for construction~\cite{Ryu_Tokyo2007}.

Concerning the Kamioka site, the Mozumi Mine, which is the current Super-Kamiokande site, cannot accomodate Hyper-Kamiokande.
A new site in Tochibora at a depth of 1400-1900 m.w.e. was found, which is 
located about 8~km south of the Mozumi Mine. This location allows for a solution to provide
the T2K neutrino beam with the same spectral properties to both Super-Kamiokande and Hyper-Kamiokande~\cite{Wakabayashi_Tokyo2007}.

\section{Conclusion}

All participants of the workshop agreed on the high physics potential of the Korean detector 
with the J-PARC beam. It was recognized that the results from the RENO and T2K experiments and
more detailed studies are required to fully optimize the advantages of a detector in this location. 
It was decided to hold a forth workshop in Korea to further explore the opportunities of this very exciting
physics program.

\section*{Acknowledgements}

The authors would like to thank all the participants at the three
International Workshop on a Far Detector in Korea for the J-PARC Neutrino 
Beam. This manuscript was prepared based on the contributions of the 
participants. 




\begin{thebibliography}{99}

\bibitem{RENO}
K.~K.~Joo {\it et al.} [RENO Collaboration],
Nucl. Phys. Proc. Suppl. {\bf 168}, 125 (2007);
see also, E.~Jeon, in these proceedings.

\bibitem{Apollonio:1999ae}
  M.~Apollonio {\it et al.}  [CHOOZ Collaboration],
  Phys.\ Lett.\ B {\bf 466}, 415 (1999)
  [arXiv:hep-ex/9907037].
 
 \bibitem{ishida:JPARC}
  The joint project team of JAERI and KEK (2000), 
  JAERI-Tech 2000-03 (2000), KEK Report 99-5 (in Japanese); 
  {\it http://j-parc.jp/index.html}

\bibitem {T2K}
Y.~Itow {\it et al.}, arXiv:hep-ex/0106019.\\
see also, 
K.~K.~Joo, K.~Nishikawa,
Talks at the 1st International Workshop on a Far Detector in 
Korea for the J-PARC Neutrino Beam, KIAS, 
Seoul, Korea, Nov. 18-19, 2005.

\bibitem{Ishida_Tokyo2007}
T.~Ishida,  in these proceedings.

  \bibitem{NP08} See session
  ``Neutrino Experiments and Proton Decay Experiments''
  (Convener T.~Hasegawa), in the
  4th International Workshop
on Nuclear and Particle Physics at J-PARC
(NP08), March 2008.

\bibitem{korea-map}
http://www2.yukawa.kyoto-u.ac.jp/\~~okamura/T2Kr/

\bibitem{Ishitsuka:2005qi}
  M.~Ishitsuka, T.~Kajita, H.~Minakata and H.~Nunokawa,
  Phys.\ Rev.\  D {\bf 72}, 033003 (2005)
  [arXiv:hep-ph/0504026].

\bibitem{Marciano:2001tz}
  W.~J.~Marciano,
  ``Extra long baseline neutrino oscillations and CP violation,''
  arXiv:hep-ph/0108181.

\bibitem{Diwan_Tokyo2007}
M.~V.~Diwan,  in these proceedings.


\bibitem{Rubbia-Meregaglia-Seoul2006} 
A.~Rubbia and A.~Meregaglia, 
Talk at the 2nd International Workshop on a Far Detector in 
Korea for the J-PARC Neutrino Beam, Seoul National University, 
Seoul, Korea, July 13-14, 2006.
 
\bibitem{Kajita_Tokyo2007}
T.~Kajita,  in these proceedings.

\bibitem{Hagiwara}
  K.~Hagiwara, N.~Okamura and K.~i.~Senda,
  Phys.\ Lett.\ B {\bf 637}, 266 (2006)
  [arXiv:hep-ph/0504061].

\bibitem{Okamura_KIAS2005} 
N.~Okamura, 
Talk at the 1st International Workshop on a Far Detector in 
Korea for the J-PARC Neutrino Beam, KIAS, 
Seoul, Korea, Nov. 18-19, 2005.

\bibitem{Dufour-Seoul2006} 
F.~Dufour, 
Talk at the 2nd International Workshop on a Far Detector in 
Korea for the J-PARC Neutrino Beam, Seoul National University, 
Seoul, Korea, July 13-14, 2006.

\bibitem{Rubbia-Seoul2006} 
A.~Rubbia, 
Talk at the 2nd International Workshop on a Far Detector in 
Korea for the J-PARC Neutrino Beam, Seoul National University, 
Seoul, Korea, July 13-14, 2006.

\bibitem{Dufour_Tokyo2007}
F.~Dufour,  in these proceedings.

\bibitem{Huber_Tokyo2007}
P. Huber, in these proceedings.
\bibitem{Okumura_Tokyo2007}
K. Okumura, in these proceedings.
\bibitem{Okamura_Tokyo2007}
N. Okamura, in these proceedings.

\bibitem{Senda_Tokyo2007}
T.~Senda,  Talk at the 3rd International Workshop 
                       on a Far Detector in Korea for the J-PARC 
                       Neutrino Beam, Univ. of Tokyo, Tokyo, Japan, Sep.
                       30 - Oct. 1, 2007.  
                       
\bibitem{Rubbia_KIAS2005}
A.~Rubbia,
Talk at the 1st International Workshop on a Far Detector in 
Korea for the J-PARC Neutrino Beam, KIAS, 
Seoul, Korea, Nov. 18-19, 2005.

\bibitem{Rubbia_Tokyo2007}
A.~Rubbia,  in these proceedings.

\bibitem{Ishihara_Tokyo2007}
 C. Ishihara, in these proceedings.

\bibitem{Ko_Tokyo2007}
P. Ko, in these proceedings.  

\bibitem{Kimura_Tokyo2007}
K. Kimura, Talk at the 3rd International Workshop 
                       on a Far Detector in Korea for the J-PARC 
                       Neutrino Beam, Univ. of Tokyo, Tokyo, Japan, Sep.
                       30 - Oct. 1, 2007.  
                       
\bibitem{Koshiba:mw}
M.~Koshiba,
``Kamioka Nucleon Decay Experiment: Kamiokande Collaboration,'' 
Lecture given at SLAC Summer Institute, July 1988.

\bibitem{Fukuda:2002uc}
Y.~Fukuda {\it et al.},
Nucl.\ Instrum.\ Meth.\ A {\bf 501}, 418 (2003).

\bibitem{Ahn:2006zza}
  M.~H.~Ahn {\it et al.}  [K2K Collaboration],
  Phys.\ Rev.\  D {\bf 74}, 072003 (2006)
  [arXiv:hep-ex/0606032].
  
\bibitem{Nakamura:2003hk}
K.~Nakamura,
Int.\ J.\ Mod.\ Phys.\ A {\bf 18} (2003) 4053.

\bibitem{t600paper}
S.~Amerio {\it et al.},
Nucl. Instrum. Meth. A 527 (2004) 329.

\bibitem{Rubbia:2004tz}
  A.~Rubbia,
  ``Experiments for CP-violation: A giant liquid argon scintillation,  Cherenkov
  and charge imaging experiment?,''
  arXiv:hep-ph/0402110.

\bibitem{Ryu_Tokyo2007}
C. Ryu, in these proceedings.

\bibitem{Wakabayashi_Tokyo2007}
N. Wakabayashi {\it et al.},  in these proceedings.

\end{thebibliography}
\end{document}